\documentstyle[12pt,epsf]{article}
\newcommand{\be}{\begin{eqnarray}}
\newcommand{\ee}{\end{eqnarray}}
\newcommand{\nn}{\nonumber}

{\newcommand{\lsim}{\mbox{\raisebox{-.6ex}{~$\stackrel{<}{\sim}$~}}}
{

\def\0n{0\nu\beta\beta}

\def\pto{\mathrel{\vcenter{\hbox{${\cal P}$}\nointerlineskip\hbox{$\rightarrow$}}}}
\def\dto{\mathrel{\vcenter{\hbox{${\cal D}$}\nointerlineskip\hbox{$\rightarrow$}}}}
\def\poto{\mathrel{\vcenter{\hbox{${\cal P}$}\nointerlineskip\hbox{$\leftrightarrow$}}}}
\def\doto{\mathrel{\vcenter{\hbox{${\cal D}$}\nointerlineskip\hbox{$\leftrightarrow$}}}}
\def\d{${\cal D}~$}
\def\p{${\cal P}~$}
\begin{document}
\thispagestyle{empty}
\begin{center}
{\LARGE \bf Parity in left-right symmetric models \\}
\vspace{1.0in}

{\bf Utpal Sarkar\\}

\vspace{0.1in}

{\sl Physical Research Laboratory, Ahmedabad 380 009, India\\}

\vspace{0.1in}
\end{center}
\vspace{.5in}
\begin{abstract}

We considered parity breaking in some left-right symmetric models.
We studied spontaneous and explicit parity violation in two cases
with doublet and triplet Higgs scalars. The minimization condition
in these two cases differ significantly. A comparative study of
these models is presented emphasizing their phenomenological
consequences.

\end{abstract}

\vspace{1cm}
\newpage
\baselineskip 18pt

\section{Introduction}

The $V-A$ structure \cite{sud} of the weak interaction and parity
violation \cite{lee-yang} suggested that the low energy gauge
group should treat the left-handed particles preferentially. This
leads to the standard model \cite{std} gauge group ${\cal G}_{std}
\equiv SU(3)_c \times SU(2)_L \times U(1)_{Y}$, in which all
right-handed particles transform trivially under $SU(2)_L$.
Although this explains the parity violation at low energy, this
does not explain the origin of the parity violation. With an
attempt to explain parity violation starting from a parity
conserved theory, the standard model group was extended to a
left-right symmetric group ${\cal G}_{LR} \equiv SU(3)_c \times
SU(2)_L \times SU(2)_R \times U(1)_{B-L}$, where the group treats
all fermions in a similar way \cite{lr}. The left-handed fermions
transform trivially under $SU(2)_R$, while all right-handed
fermions transform trivially under $SU(2)_L$. One may then define
a left-right parity, which interchanges these two groups $SU(2)_L
\Leftrightarrow SU(2)_R$. Since one interchanges all left-handed
fermions with the right-handed fermions under parity, this becomes
the parity of the Lorentz group, associated with space reflection.

It is then possible to break parity spontaneously or explicitly in
left-right symmetric models. When the right-handed symmetry group
$SU(2)_R$ is broken, parity breaks down. One may also break parity
even before breaking the group $SU(2)_R$ by giving a vacuum
expectation value to a Higgs scalar, which changes sign when the
two $SU(2)$ groups are interchanged. These spontaneous breaking of
parity is quite natural in grand unified theories. However, in
some superstring inspired models it is also possible that the
parity breaks down during compactification and one ends up with
explicitly broken parity in left-right symmetric model.

Here we shall study the consequences of breaking parity in some
left-right symmetric models with triplet and doublet Higgs
scalars. Spontaneous parity breaking in models with triplet Higgs
was studied extensively \cite{dpar}. Here we discuss the models
with doublet Higgs and explicit parity breaking. We study some
aspects of these models after explaining the question of parity in
some more details. The minimization conditions differs in these
two cases with triplet and doublet Higgs scalars and imply
different low energy phenomenology.

\section{Parity of Lorentz group and left-right symmetric models}

To define the parity, consider the local Lorentz group $SO(4)$
which is isomorphic to $SU(2) \times SU(2)$. There are
inequivalent embeddings of the two $SU(2)$ in the $SO(4)$, in one
of these embeddings the two $SU(2)$ groups act on the left-handed
and right-handed fermions. Since the two $SU(2)$ groups are
indistinguishable, there is a $Z_2$ symmetry, which interchanges
the two $SU(2)$ groups. This is the usual discrete parity symmetry
or the space-reflection symmetry \p.

A scalar [pseudo-scalar] field $\phi(x)$ is even [odd] under
parity and transform under \p as
$$ \phi(x) \pto \pm \phi(-x) .$$
Similarly, a vector [pseudo-vector] field is even [odd] under \p.
However, for fermions we have to consider their chiral
decomposition. A left-handed fermion $\psi_L = {1 \over 2} (1 -
\gamma_5) \psi$ transforms under the Lorentz group $SU(2)_L \times
SU(2)$ as [2,0], while a right-handed fermion $\psi_R = {1 \over
2} (1 + \gamma_5) \psi$ transforms as [0,2]. Under \p parity they
transform as
\begin{equation}
\psi_{L} \pto \psi_R ~~~~ {\rm and} ~~~~ \psi_R \pto \psi_L.
\end{equation}
Parity does not act on any internal group space. In the standard
model, the left-handed fields are doublets under $SU(2)_L$, while
all right-handed fields are singlets. So, \p parity does not
commute with the group $SU(2)_L$ and \p is maximally broken.

In the left-right extension of the standard model, the gauge group
is ${\cal G}_{LR} \equiv SU(3)_c \times SU(2)_L \times SU(2)_R
\times U(1)_{B-L}$. Under ${\cal G}_{LR}$ the left-handed and the
right-handed quarks and leptons transform as
\begin{eqnarray}
q_L = \pmatrix{u_L \cr d_L} \equiv [3,2,1,1/3] ~~~&&~~~
q_R \pmatrix{u_R \cr d_R} \equiv [3,1,2,1/3] \nonumber \\
\ell_L \pmatrix{\nu_L \cr e_L} \equiv [1,2,1,-1] ~~~&&~~~ \ell_R
\pmatrix{\nu_R \cr e_R} \equiv [1,1,2,-1]. \nonumber
\end{eqnarray}
The electric charge $Q$ and the hypercharge $Y$ are defined by
\begin{equation}
Q = T_{3L} + T_{3R} + {B-L \over 2} = T_{3L} + Y .
\end{equation}
The $SU(2)_L$ and the $SU(2)_R$ groups are now related by a
discrete $Z_2$ symmetry, called the \d parity. Under \d the gauge
groups transform as $$SU(2)_L \doto SU(2)_R$$ and the fermions
transform as $$ \psi_L \doto \psi_R .$$ Thus the fermions cannot
distinguish between the \d and the \p parity.

Let us now consider the gauge bosons. The $SU(2)_L$ gauge bosons
$W_L$ and the $SU(2)_R$ gauge bosons $W_R$ are not parity
eigenstates. The linear combinations $$ V = W_L + W_R ~~~~~ {\rm
and} ~~~~~ A = W_L - W_R $$ have definite parities: $$ V \poto V
~~~~~ {\rm and} ~~~~~ A \poto -A .$$ This will imply that under
parity, $$ W_L \leftrightarrow W_R $$ so that we can identify the
left-right parity with the Lorentz parity. We shall now consider a
couple of left-right symmetric models with triplet and doublet
Higgs scalars and study the consequences of spontaneous and
explicit parity breaking.

\section{Left-right symmetry with triplet Higgs}

In the left-right symmetric models, a bi-doublet Higgs scalar
$$\Phi \equiv [1,2,2,0]$$ gives masses to all fermions and breaks
the electroweak symmetry. This acquires a vacuum expectation value
of 246 GeV. Then the neutrinos will also become as heavy as the
other fermions, unless the Yukawa coupling for the neutrinos are
too small which is unnatural. This problem is solved in these
models, when one includes a new triplet Higgs scalar to break the
left-right symmetry at a large scale $M_R$. In the left-right
symmetric models the scale $M_R$ is same as the scale of lepton
number violation. So, one right-handed triplet Higgs with $B-L =
2$ can break the $SU(2)_R \times U(1)_{B-L}$ symmetry and give
Majorana masses to the right-handed neutrinos, which in turn gives
a tiny see-saw mass to the left-handed neutrinos \cite{mm}.

In the simplest left-right symmetric model, one assumes parity to
be conserved before the left-right symmetry (${\cal G}_{LR}$)
breaking. The group ${\cal G}_{LR}$ is broken when the
right-handed triplet Higgs scalar ($\Delta_R \equiv [1,1,3,-2]$)
acquires a vacuum expectation value ($vev$) $\langle \Delta_R
\rangle = u_R$. The $vev$ of this right-handed triplet Higgs
scalar breaks $B-L$ by 2 units and hence it can give a Majorana
mass to the right-handed neutrinos. Parity then dictates that
there should also be a left-handed triplet Higgs scalar $\Delta_L
\equiv [1,3,1,-2]$. However, the $vev$ of $\langle \Delta_L
\rangle = u_L$ should be very small (less than a GeV) for the
theory to be consistent with $Z$ decay width. Since $u_L$ can also
give neutrino masses, it should be less than a few eV.

In general, the masses ($\mu_L$ and $\mu_R$) of $\Delta_L$ and
$\Delta_R$ could be different, breaking parity explicitly.
However, when parity is conserved these masses are same
($\mu_L=\mu_R$). In models of spontaneous parity breaking, one
introduces an additional singlet Higgs scalar ($\sigma \equiv
[1,1,1,0]$, with $\langle \sigma \rangle = \eta$), which is odd
under \d parity:
$$\sigma \dto - \sigma.$$ For a general analysis of the scalar
potential, we shall include the scalar $\sigma$ and also allow
$\mu_L \neq \mu_R$. The parity conserving scenario, spontaneous
parity breaking scenario and the explicit parity breaking scenario
will then come out as special cases of this analysis.

The scalar potential will contain several terms including all
these fields: the bi-doublet $\Phi$, the right-handed triplet
$\Delta_R$, the left-handed triplet $\Delta_L$ and the odd \d
parity singlet $\sigma$. To study the minima of this potential we
replace these fields by their $vev$ without loss of generality. We
write the potential in several components:
\begin{equation}
V(\kappa,u_L,u_R,\eta) = V_\kappa + V_{L} + V_{R} + V_\eta +
V_{\kappa LR} + V_{\eta LR} + V_{\kappa \eta} + V_{LR}
\end{equation}
where $\langle \Phi \rangle = \pmatrix{0& \kappa \cr \kappa^\prime
& 0}$ with $\kappa \gg \kappa^\prime$. The first few terms contain
the usual quadratic and quartic couplings
\begin{equation}
V_a = - {1 \over 2} \mu_a^2 \phi_a^2 + {1 \over 4} \lambda_a
\phi_a^4
\end{equation}
where $a = \kappa,L,R,\eta$ and $\phi_a = \kappa,u_L,u_R,\eta$,
respectively. The cross terms are:
\begin{eqnarray}
V_{\kappa LR} &=& {1 \over 2} \left( g_L u_L^2 + g_R u_R^2 + 2
g_{LR} u_L u_R \right) \kappa^2 \nn \\
V_{\eta LR} & = & {1 \over 2} M \eta ( u_L^2 - u_R^2) + {1 \over
2} \lambda_2 \eta^2 ( u_L^2 + u_R^2) \nn \\
V_{\kappa \eta} &=& \bar M \eta \kappa^2 + {1 \over 2} \lambda_1
\eta^2 \kappa^2 \nn \\
V_{LR} &=& {1 \over 2} g u_L^2 u_R^2
\end{eqnarray}
The term with coefficient $g_{LR}$ plays a very significant role,
as we shall discuss later.

\subsection*{$SU(2)_R$ and parity}

We shall first consider the most popular model in which parity is
broken when the group $SU(2)_R$ is broken. In this case, $\eta =
0$ signifies there is no spontaneous symmetry breaking.
Conservation of the \d parity will constrain some of the
parameters:
\begin{equation}
\mu_L = \mu_R = \mu_\Delta, ~~~~~ g_L = g_R = g^\prime ~~~~~
\lambda_L = \lambda_R = \lambda_\Delta . \label{cond}
\end{equation}
Minimization of the potential then gives a condition:
\begin{equation}
u_L {\partial V \over \partial u_R } - u_R {\partial V \over
\partial u_L } = 0 = (u_L^2 - u_R^2) [ g_{LR} \kappa^2 + (g -
\lambda_\Delta) u_L u_R ].
\end{equation}
One of the solution $u_L = u_R$ preserves parity and left-right
symmetry is not broken, which is not acceptable
phenomenologically. The other solution breaks parity and
determines the $vev$ of the left-handed triplet Higgs scalar
$\Delta_L$ to be,
\begin{equation}
u_L = {g_{LR} \kappa^2 \over (\lambda_\Delta -g) u_R }.
\end{equation}
The smallness of the $vev$ of the left-handed triplet Higgs scalar
is thus guaranteed in the limit of large $u_R$ {\it i.e.}, a large
lepton number violating scale. The term with coefficient $g_{LR}$
is very crucial for this condition. After $\Delta_R$ acquires a
$vev$, this term allows an interaction of the left-handed triplet
Higgs with two bi-doublets. This term and the Yukawa coupling of
the triplet Higgs with the leptons together then breaks lepton
number. Thus the see-saw suppressed $vev$ of the triplet Higgs and
the lepton number violation is intimately related \cite{trip}.

Although the $vev$ of the left-handed triplet Higgs scalar is
tiny, the masses of both the triplet Higgs scalars are very large.
For a neutrino mass of less than eV (as required by present
experiments), the scale of lepton number violation is required to
be greater than $10^7$ GeV. Thus these Higgs scalars will not be
accessible to the next generation accelerators.

\subsection*{Spontaneous parity violation}

For spontaneous parity violation \cite{dpar}, we start with a
parity conserving model, so that the constraints on the couplings,
given by equation \ref{cond} is satisfied. We then consider $\eta
\neq 0$. The minimization of the potential then gives the
condition:
\begin{equation}
u_L {\partial V \over \partial u_R } - u_R {\partial V \over
\partial u_L } = 0 = 2 M \eta u_L u_R +
(u_L^2 - u_R^2) [ g_{LR} \kappa^2 + (g - \lambda_\Delta) u_L u_R
].
\end{equation}
In this case, $g_L = g_R$ is no longer a solution of the
minimization condition, except for $\eta =0$.

All the solutions of this minimization condition breaks parity.
For simplification we consider the solution of our interest, {\it
i.e.}, $u_L \ll u_R$. This leads to a $vev$ of the left-handed
triplet Higgs scalar
\begin{equation}
u_L = {g_{LR} u_R \kappa^2 \over (\lambda_\Delta -g) u_R^2 - 2 M
\eta }.
\end{equation}
The scale of parity violation $\eta$ now determines the $vev$ of
the triplet Higgs. Thus, the left-right symmetry breaking scale
can be much lower. The mass of the triplet Higgs can also be
small, which can then be detected in the next generation
accelerators.

\subsection*{Explicit parity violation}

Since parity is violated explicitly, we do not require the scalar
$\sigma$. However, if we include this field in our analysis, this
will not change anything. For comparison with the spontaneous
parity breaking scenario, we shall retain both these
contributions. The minimization condition now reads
\begin{eqnarray}
u_L {\partial V \over \partial u_R } - u_R {\partial V \over
\partial u_L } = 0 &= & u_L u_R [ (\mu_L^2 -\mu_R^2) + 2 M \eta
+ (g_L - g_R) \kappa^2 \nn \\ &+& (\lambda_L u_L^2 - \lambda_R
u_R^2) ] + (u_L^2 - u_R^2) [ g_{LR} \kappa^2 + g  u_L u_R ]. \nn
\\ &&
\end{eqnarray}
The explicit breaking contribution $\mu_L^2 - \mu_R^2$ can now
substitute for the spontaneous symmetry breaking term $2 M \eta$.
In this scenario also, there is no solution with $g_L = g_R$. The
parity violating solution correspond to
\begin{equation}
u_L = {g_{LR} u_R \kappa^2 \over (\mu_L^2 -\mu_R^2) + 2 M \eta -
\lambda_R u_R^2 + g u_R^2 }.
\end{equation}
In this expression we neglected $\kappa^2$ and $u_L^2$ terms
compared to $u_R^2$ and $\eta^2$, assuming there is no fine
tuning.

There is no distinction between the spontaneously broken parity
and explicit breaking. In grand unified theories spontaneously
broken \d parity is quite natural, while in superstring inspired
models the \d parity can be broken by Wilson loops at the time of
compactification. In these theories explicit breaking of parity
appear to be more natural. A light triplet Higgs with rich
phenomenology is allowed even with explicit parity violation.

\section{Left-right symmetry with doublet Higgs}

The left-right symmetry may also be broken by a doublet Higgs
scalar \cite{doub}. For charged fermion masses one may retain a
bi-doublet scalar $\Phi$, which also breaks the electroweak
symmetry. Although it is possible to construct models without any
bi-doublet, here we restrict ourselves to models including the
bi-doublet. The $SU(2)_R \times U(1)_{B-L}$ group is broken by a
right-handed Higgs doublet $\chi_R \equiv [1,1,2,+1]$, which
acquires a $vev$ $\langle \chi_R \rangle = v_R$. Since the $vev$
of this doublet breaks lepton number by 1 unit, it cannot
contribute to the neutrino masses. However, including a singlet
heavy fermion this problem can be solved, which we shall discuss
later.

The left-right parity then ensures that there is another
left-handed doublet Higgs scalar field $\chi_L \equiv [1,2,1,+1]$,
which acquires a small $vev$ $\langle \chi_L \rangle = v_L$.
Unlike the left-handed triplet Higgs $vev$ $u_L$, there is no
restriction on the $vev$ of the doublet Higgs $v_L$ except that
$v_L < \kappa$, since it breaks the group $SU(2)_L$.

The scalars $\chi_{L,R}$ break lepton number by one unit, so they
cannot give masses to the neutrinos. Let us now consider the
effective dimension-5 terms, which can contribute to the neutrino
masses,
\begin{equation}
{\cal L}_\nu = {h_L \over M_L} \ell_L \ell_L \chi_L \chi_L + {h_R
\over M_L} \ell_R \ell_R \chi_R \chi_R .
\end{equation}
$M_L$ is the lepton number violating scale. Since $B-L$ is broken
at a scale $v_R$, we expect $M_L \sim v_R$. There are no terms
involving the bi-doublet $\Phi$, since it does not carry any $B-L$
quantum number.

For a realization of this effective operator, we introduce a heavy
singlet fermion $S \equiv [1,1,1,0]$. This field $S$ will interact
with the neutrinos and the Higgs scalars and will affect the
neutrino masses. We write down the terms which contribute to the
neutrino masses
\begin{equation}
{\cal L}_S = f_{iL} \bar \ell_{iL} S \chi_L + f_{aR} \bar
\ell_{aR} S \chi_R + M_S S S + f_{ia} \bar \ell_L \ell_R \Phi ,
\end{equation}
where $i = 1,2,3$ represents the three generations of left-handed
leptons and $a=1,2,3$  represents the three generations of
right-handed leptons. When the scalars acquire $vev$s, they
contribute to the neutrino mass matrix. We write down the neutrino
mass matrix in the basis $\pmatrix{\nu_{iL} & \nu_{aR} & S}$
\begin{equation}
M_\nu = \pmatrix { 0 & f_{ia} \kappa & f_{iL} v_L \cr f_{ia}
\kappa & 0 & f_{aR} v_R \cr f_{iL} v_L & f_{aR} v_R & M_S } .
\end{equation}
The singlet $S$ is the heaviest fermion with mass $M_S$. The
right-handed neutrinos get an effective Majorana mass of
$v_R^2/M_S$. The left-handed neutrinos get two contributions to
their mass, a see-saw contribution from the left-handed doublets
and a double-see-saw contribution from the right-handed doublets.
They are,
\begin{equation}
m_{\nu i j} = - {f_{iL} f_{j L} v_L^2 \over M_S } + {f_{i a}
f_{jb} \over f_{aR} f_{bR}} {M_S \kappa^2 \over v_R^2} .
\end{equation}
In the triplet Higgs scenario $u_L$ is too small and hence the
first term could be negligible. But as we shall see later, $v_L$
could be comparable to $\kappa$ and hence could dominate over the
second term.

To break the parity we introduce a field $\sigma \equiv [1,1,1,0]$
(with $\langle \sigma \rangle = \eta$), which is odd under \d. The
scalar potential contains the usual quadratic and quartic terms
and in addition there are some cross terms given by,
\begin{eqnarray}
V_{\chi} &=& {1 \over 2} \left( g_L v_L^2 + g_R v_R^2
\right) \kappa^2 + \mu v_L v_R \kappa   + {1 \over 2} g v_L^2 v_R^2 \nn \\
&  & + {1 \over 2} \eta [ \bar M \kappa^2 + M ( v_L^2 - v_R^2) ] +
{1 \over 2} \eta^2 [\lambda_1 \kappa^2 + \lambda_2 ( v_L^2 +
v_R^2)]
\end{eqnarray}
This potential is somewhat similar to the triplet Higgs scenario,
but still there are some important differences.

\subsection*{$SU(2)_R$ and parity}

Conservation of parity would require $\eta=0$ and
\begin{equation}
\mu_L = \mu_R = \mu_\chi, ~~~~~ g_L = g_R = g^\prime ~~~~~
\lambda_L = \lambda_R = \lambda_\chi .
\end{equation}
With these conditions we minimize the potential to obtain a
condition relating the different $vev$s given by
\begin{equation}
v_L {\partial V \over \partial v_R } - v_R {\partial V \over
\partial v_L } = 0 = (v_L^2 - v_R^2) [ \mu \kappa + (g -
\lambda_\chi) v_L v_R ].
\end{equation}
One of the solutions $v_L = v_R$ conserves left-right symmetry
even after symmetry breaking, which is of no interest. The other
solution determines the $vev$ of the left-handed triplet Higgs
\begin{equation}
v_L = { \mu \kappa \over ( \lambda_\Delta - g ) v_R }.
\end{equation}
A natural assumption for the mass scales is $\mu \sim v_R$ and
hence $v_L \lsim \kappa$, since $\kappa$ breaks the electroweak
symmetry and give masses to the fermions. The masses of both the
doublets $\chi_{L,R}$ are large $\mu_\chi \sim v_R$. Thus this
scenario does not lead to any interesting phenomenology.

\subsection*{Spontaneous parity violation}

Here we assume the \d parity conserving constraints on the
coupling constants as mentioned above, but $\eta \sim v_R$. This
would then give us the minimization condition
\begin{equation}
v_L {\partial V \over \partial v_R } - v_R {\partial V \over
\partial v_L } = 0 = 2 M \eta v_L v_R +
(v_L^2 - v_R^2) [ \mu \kappa + (g - \lambda_\chi) v_L v_R ].
\end{equation}
Similar to the triplet Higgs models, $g_L = g_R$ is no longer a
solution of the minimization condition. Assuming, $v_L \ll v_R$ we
obtain the $vev$ of the doublet Higgs to be
\begin{equation}
v_L = {\mu v_R \kappa \over (\lambda_\chi - g ) v_R^2- 2 M \eta}.
\end{equation}
The $vev$ of the doublet Higgs can again be of the order of
electroweak symmetry breaking scale. In this case the mass of the
left-handed doublet Higgs scalar can also be as small as the
electroweak symmetry breaking scale. This will make this scenario
phenomenologically interesting.

\subsection*{Explicit parity violation}

We take both $\eta \neq 0$ and $\mu_L \neq \mu_R$. For simplicity,
we shall impose \d parity on other couplings, although relaxing
this condition will not change any of the conclusions. The
minimization condition now reads
\begin{equation}
v_L {\partial V \over \partial v_R } - v_R {\partial V \over
\partial v_L } = 0 =  v_L v_R [ (\mu_L^2 -\mu_R^2) + 2 M \eta]
+ (v_L^2 - v_R^2) [\mu \kappa + (\lambda_\chi -g) v_L v_R ].
\end{equation}
It is apparent that the explicit parity breaking contribution
$\mu_L^2 - \mu_R^2$ is similar to the spontaneous parity breaking
contribution $2 M \eta$. The $vev$ of the left-handed doublet is
now given by
\begin{equation}
v_L = {- \mu v_R \kappa \over (\mu_L^2 -\mu_R^2) + 2 M \eta +(g-
\lambda_\chi) v_R^2  }.
\end{equation}
This result is similar to the spontaneous parity breaking
scenario, except that now we can decouple the left-handed and
right-handed doublets by considering $\mu=0$. The $SU(2)_R$
symmetry breaking can take place with a Higgs doublet $\chi_R$ at
a very high scale. Since parity is explicitly broken, the
left-handed doublet $\chi_L$ can now be light and can have a $vev$
$v_L \sim \mu_L$. No fine tuning is required to maintain this
solution. The light Higgs doublet will then contribute to the low
energy phenomenology.

\section{Summary}

We discussed the question of parity violation in left-right
symmetric model. The left-right parity, acting in the left-right
gauge group space, can be identified with the usual Lorentz parity
or the space reflection symmetry. Since the left-right parity is
identified as the space reflection parity, one may construct
left-right symmetric model in which parity is conserved initially
and broken along with $SU(2)_R$ symmetry breaking. Parity could
also be spontaneously broken at a different scale compared to the
left-right symmetry breaking. We also studied the explicit parity
breaking in a couple of left-right symmetric models. Depending on
whether the Higgs field required for the $SU(2)_R$ breaking is a
triplet or a doublet, the minimization of the potential and
phenomenological consequences are different.

\subsubsection*{Acknowledgement}

I would like to thank Prof. R.N. Mohapatra and Prof. R. Foot for
correcting one serious error in the first version of this
manuscript.

\end{document}